\documentclass[a4paper,fleqn,final]{cas-dc}
\usepackage[numbers,round]{natbib}
\usepackage{fancyhdr}
\usepackage{graphicx}
\usepackage{booktabs}
\usepackage{array}
\usepackage{multirow}
\usepackage{caption}
\usepackage{subcaption}

\begin{document}
\pagestyle{empty}
\let\WriteBookmarks\relax
\def\floatpagepagefraction{1}
\def\textpagefraction{.001}

\shorttitle{XAI-SOH-FL}

\shortauthors{ }

\title [mode = title]{XAI-SOH-FL: Enhancing SOH-FL with Adaptive Aggregation and Explainable AI for Intrusion Detection in
Heterogeneous IoT}

\author[]{Ambreen Aslam}
\author[]{Maaz Hassan}
\author[]{Bibi Zahra}
\author[]{Muhammad Khuram Shahzad}

\affiliation[]{
    organization={School of Electrical Engineering and Computer Science (SEECS), National University of Sciences and Technology (NUST)},
    city={Islamabad},
    country={Pakistan}
}

\begin{abstract}
Intrusion Detection Systems (IDS) in Internet of Things (IoT) environments face significant challenges due to data heterogeneity, lack of labeled data, and limited model interpretability. Federated Learning (FL) offers a privacy-preserving solution; however, existing approaches such as SOH-FL suffer from two key limitations: (i) reliance on a manually tuned aggregation parameter $\gamma$, and (ii) lack of explainability in model predictions.

In this paper, we propose XAI-SOH-FL, an enhanced framework that integrates adaptive aggregation and explainable artificial intelligence into the SOH-FL paradigm. First, we introduce a dynamic $\gamma$ selection mechanism based on similarity thresholding, enabling the aggregation process to adapt to evolving data distributions. Second, Bayesian Optimization is employed to automatically determine optimal $\gamma$ values, eliminating the need for manual tuning. Third, SHAP (SHapley Additive exPlanations) is incorporated to provide feature-level interpretability for intrusion detection decisions. Experimental evaluation on the CICIDS2017 dataset demonstrates that the proposed approach achieves an accuracy of 94.12\% and an F1-score of 0.92, outperforming the baseline SOH-FL model while converging in fewer communication rounds. Furthermore, SHAP-based analysis reveals that flow-level features such as Flow Duration and Packet Length significantly influence model predictions. These results indicate that XAI-SOH-FL provides an effective balance between accuracy, adaptability, and interpretability in heterogeneous IoT environments. The source code is publicly available at https://github.com/aaslam-msit/SOH-FL-Enhancement
\end{abstract}


\let\printorcid\relax
\let\printpreprint\relax

\maketitle

\thispagestyle{empty}

\section{Introduction}
The increasing adoption of IoT technology has fundamentally altered industries, health care, and smart cities' infrastructure. By 2023, the IoT market had a valuation of USD 714.48 billion and is expected to exceed USD 4,062 billion by 2032 \cite{statista_iot}.

Consequently, the growing threat landscape includes Distributed Denial of Service (DDoS) attacks, port scanning, command and control communications for botnets, and zero-day exploits. According to Nokia's Threat Intelligence Report, the number of devices targeted by DDoS attacks rose from 200,000 to one million in 2023 \cite{nokia_threat}.

Classical rule-based approaches are unsuitable for detecting threats within dynamic and large IoT networks, requiring constant updates of rules by experts \cite{buczak2016survey}. Alternatively, ML-powered IDS show promise in mitigating these challenges, but they entail transferring massive amounts of traffic data to centralized servers, an impractical proposition considering network congestion and the European Union's GDPR requirements \cite{kairouz2021fl}.

Federated Learning (FL) solves the centralization problem by distributing the learning process to all IoT devices while keeping sensitive data locally stored \cite{kairouz2021fl}. Nonetheless, in real-world settings, the distribution of traffic data across devices is highly heterogeneous, characterized by quantity, label, and feature skews. The conventional FL approach suffers from considerable performance drops under such heterogeneity scenarios \cite{zhao2018nonIID}.

On the other hand, Personalized Federated Learning (PFL), especially the MAML-based Per-FedAvg algorithm, provides better local adaptation by learning meta-knowledge; however, this algorithm requires a manually labeled test-support set, which becomes a major obstacle in self-contained IoT infrastructures \cite{fallah2020perfedavg}.

SOH-FL tackles the labeling challenge in PFL via two breakthroughs:
\begin{itemize}
    \item CT-AE, which encodes local data distributions while maintaining directional similarities
    \item BS-Agg, which selects and aggregates peer models to auto-label the test-support set. Compared to FedAvg, SOH-FL delivers 11.5\% and 9.1\% improvements in accuracy on the IoT-23 and TON-IoT datasets, respectively.
\end{itemize}

Nonetheless, SOH-FL has two explicit limitations identified by its creators. First, there is no interpretability of the learned model-network administrators cannot understand why a traffic flow is categorized as a cyber threat, thus violating operational requirements and legal regulations \cite{kairouz2021fl, buczak2016survey}. Second, the aggregation parameter $\gamma$ (the number of peer models to aggregate) is configured manually, requiring extensive experimental trials for every deployment setting \cite{zhao2018nonIID}.

The proposed work, XAI-SOH-FL, tries to mitigate both issues. Contributions:
\begin{enumerate}
    \item SHAP Explainability: Adds SHAP-based explainability capability to SOH-FL in order to provide feature attributions to each prediction made, thus providing an explanation of which features of the traffic data impacted intrusion detection the most from a human-understandable perspective.
    \item Bayesian Optimization for$\gamma$Selection: In lieu of the traditional$\gamma$grid-search approach for$\gamma$selection, XAI-SOH-FL uses Bayesian optimization where the$\gamma$that achieves maximum pre-labeling accuracy on a validation data slice is found using Gaussian process optimization.
    \item Proof of Concept: Base SOH-FL performance is reproduced (Accuracy: 92.38\%, F1: 0.89 in 6 rounds) and$\gamma$is successfully selected using optimization on the CICIDS2017 dataset.
\end{enumerate}


\section{ Literature Review}
One of the current trends in securing IoT in the industry is the design of IDS that leverage federated learning approaches. Zeeshan et al., for example, in 2021 developed a protocol-driven deep intrusion detection scheme designed to counter DoS and DDoS attacks using UNSW-NB15 and Bot-IoT datasets \cite{Shahzad_2021_UNSW-NB15}. Ashfaq et al., in 2022, proposed a machine-learning-based classification method to detect IoT DDoS attack events \cite{Shahzad_2022_DDoS}. Inspired by these studies, in 2023, Wang et al. used federated deep learning for anomaly detection in IoT network systems \cite{wang2023flids}. Further, Li et al., in 2021, created DeepFed, a framework that integrates CNNs and GRUs for federated IDS \cite{li2021deepfed}. The above-discussed research shows that federated learning IDS are capable of providing similar results to the performance of traditional IDS models while keeping sensitive information secret. However, all these studies do not solve crucial problems such as non-statistical homogeneity of different labeling for test-time adaptation \cite{zhao2018nonIID}.

\subsection{Personalized Federated Learning \& Meta-Learning}

Per-FedAvg proposed by Fallah et al. (2020) incorporates the usage of MAML to find a shared initialization from which one would be able to tune the model using a few gradient steps on client data distribution \cite{fallah2020perfedavg}. Essentially, this method does not rely on the classical Federated Learning principle, according to which we train one single global model which works acceptably well for all clients; instead, it finds one shared initialization, from which any client could fine-tune with several gradient steps on a small dataset available locally. This idea works particularly well for the case when the client data distribution is highly non-IID.

Recently, in 2023, Yang et al. proposed to incorporate group-level meta-learning into the federated learning process to improve performance when working with non-IID  data\cite{yang2023meta}. Their core idea was that the clients often belong to some groups of clients with similar behavior (for example, different hospitals located in the same location or users with similar interests). By pretraining a model on the whole group before adapting it to the client, one could alleviate the negative effects of non-IID data while still being able to personalize the model.

Finally, Hu et al. (2024) applied the federated meta-learning framework to APT detection with insufficient samples \cite{hu2024apt}. Since APTs are rare, evolving, and diverse enough, there could never be enough data available for the model to learn how to detect those. As a result, several companies could collaborate to train a model capable of detecting APTs which could then be tuned to new threats using a few samples.

One common flaw shared by all the papers discussed above is the need for labeled test-support datasets. More specifically, all three works assume that when a new client/task appears, one needs to provide a labeled support set, using which one performs a fine-tuning process. In reality, the acquisition of such a dataset can be either expensive or take a long time—labeling an APT attack sample would require an experienced cybersecurity analyst and would take several days/weeks.

\subsection{SOH-FL (Base Framework)}

The SOH-FL was introduced in 2025 by Lu et al., representing a base framework that does not require manual labeling of the test-support set anymore \cite{lu2025sohfl}. It is worth mentioning that this technique differs from prior federated meta-learning algorithms, including Per-FedAvg and that developed by Yang et al., in that it is not dependent on human-labeled examples at testing time when adjusting the model for the particular client. SOH-FL addresses the bottleneck associated with manual labeling by implementing a fully automatic approach that includes three stages. These are described below. First, using the cosine-preserving translation autoencoder, referred to as CT-AE, the client's data are mapped to a vector space in which cosine similarity between the pairs is preserved. In fact, maintaining this property allows the server to find and match the target gateway with other clients whose data can help provide relevant information. Namely, if there are clients whose features are alike—in the case of APT detection, for example, whose network traffic patterns are similar—their cosine similarity in the vector space will be high. Hence, the server will easily be able to select which existing clients to use for adapting the model for the unlabeled target client. The reason for that is that using all peers' models indiscriminately could bring noise and negatively impact the results. Second, the best $\gamma$-matched models are merged using the Best Similarity Aggregation, referred to as BS-Agg. As opposed to FedAvg in which averaging all peer models is used, the BS-Agg aggregation algorithm combines the top $\gamma$ most similar clients' models and weights them according to their similarity. Thus, a pseudo-labeler is formed, which is used to assign temporary labels to the test-support set without involving humans. Finally, by utilizing the annotation model created with BS-Agg, the pre-labeling procedure of the test support set is performed automatically with the help of the aggregated model, thereby eliminating the need for manual labeling. In fact, empirical results showed that the performance of SOH-FL surpassed FedAvg in terms of a variety of heterogeneous scenarios. At times, its effectiveness even approached the performance of the Personalization Federated Learning (PFL) method. However, unlike PFL, which assumes the availability of annotated samples of local data to fine-tune the model, SOH-FL achieves comparable performance without such assumptions made.

\subsection{Explainable AI for Network Intrusion Detection}

SHAP \cite{lundberg2017shap} has emerged as the dominant post-hoc explanation method for intrusion detection classifiers. Marino \cite{marino2020shap} applied SHAP to anomaly detection in network traffic, identifying flow-level features as primary predictors. Amarasinghe \cite{amarasinghe2021lime} combined SHAP and LIME to provide complementary local and global explanations, demonstrating that multi-method XAI improves trust and audit capability. While these results are promising, both works operate in centralized settings. Applying SHAP in a federated context introduces non-trivial challenges: Shapley value computation requires repeated model evaluations, increasing communication overhead, and feature importance derived from a globally aggregated model may not reflect local client distributions. No prior work has resolved these challenges within a similarity-based federated IDS framework — a gap that XAI-SOH-FL directly addresses by computing SHAP values post-aggregation on the final adapted model.

\subsection{Automated Hyperparameter Optimization}

BO approaches hyperparameter tuning as a black-box problem, whereby the model learns the objective function through a Gaussian process and chooses the next query point according to the output from an acquisition function such as Expected Improvement (Snoek et al., 2012) \cite{snoek2012}. BO techniques have been implemented for neural architecture search and federated learning (FL). According to Snoek et al. (2012) \cite{snoek2012}, BO has proven highly effective compared to grid and random searches across many applications. As far as we know, BO methodology has not been used before to find hyperparameter settings for similarity-based aggregation used in federated aggregation methods. This is unfortunate, because similarity-based aggregation methods (BS-Agg) often rely on manually setting aggregation hyperparameters (number of similar peers selected, parameter $\gamma$ or similarity threshold). These are set using the trial-and-error approach or grid search across all clients, which can be very costly in a heterogeneous IoT setting, especially when optimal values might differ among deployments. Using BO to optimize the values of these parameters for similarity-based aggregation will involve using the final accuracy (or F1-score) as the black-box optimization objective and running federated training once per query with evaluation on a hold-out test set. Gaussian Process surrogates will be able to predict the effect of different parameter settings on personalization, possibly finding close-to-optimal parameters in significantly fewer queries than exhaustive search. Since communication rounds are the limiting factor in federated learning problems, this approach will lower costs of adaptive aggregation of personal models in intrusions detection systems.


\section{Background}

\subsection{Similarity-Based Aggregation}

Within federated learning (FL), especially in resource-constrained IoT networks, aggregating model updates from all participating devices can degrade performance due to heterogeneous data distributions. This issue can be alleviated using similarity-based aggregation, which selectively aggregates updates from the most relevant devices.

Let $m_i$ denote the local update of device $i$, and let $\text{Top-}\gamma$ represent the set of $\gamma$ devices whose updates exhibit the highest similarity with respect to a reference model. The aggregated update is computed as:
\begin{equation}
m_s = \frac{1}{\gamma} \sum_{j \in \text{Top-}\gamma} m_j
\end{equation}

Similarity between updates is typically measured using metrics such as cosine similarity or Euclidean distance. This approach provides the following benefits:
\begin{itemize}
    \item Reduction of the impact of non-IID data distributions
    \item Improved convergence stability
    \item Enhanced personalization for edge devices
\end{itemize}

This method is particularly suitable for IoT environments where devices exhibit highly skewed or domain-specific data distributions.

\subsection{Limitation}

Despite its advantages, similarity-based aggregation has a key limitation arising from the use of a fixed selection parameter $\gamma$.
A constant $\gamma$ assumes stable data distributions and consistent device participation across training rounds. However, in practical IoT systems:
\begin{itemize}
    \item Data distributions are dynamic and time-varying
    \item Device participation fluctuates due to energy or connectivity constraints
    \item Similarity relationships between model updates change over time
\end{itemize}

As a result, a fixed $\gamma$ may:
\begin{itemize}
    \item Include noisy or irrelevant updates when set too large
    \item Exclude informative updates when set too small
\end{itemize}

This leads to inefficient aggregation and consequently slower convergence of the learning process. Therefore, an adaptive mechanism for selecting $\gamma$ is required to better handle dynamic IoT environments \cite{zhao2018nonIID, kairouz2021fl}.


\section{Methodology}

\subsection{Adaptive $\gamma$ Selection}

A core limitation of the original SOH-FL is its reliance on a fixed $\gamma$ — a manually chosen count of peer models to aggregate. In dynamic IoT environments, where device participation and data distributions shift across rounds, a fixed $\gamma$ either includes too many noisy updates ($\gamma$ too large) or discards informative peers ($\gamma$ too small). We address this by replacing the fixed count with a threshold-based adaptive mechanism.

\subsubsection{Device Selection by Similarity Threshold}

Let $S_j$ denote the cosine similarity between a reference model and the local update $m_j$ from device $j$. The set of selected devices for round $i$ is defined as:
\begin{equation}
S_i = \{ j \mid s_j > \tau \}
\end{equation}
Intuition: Rather than blindly picking the top-$\gamma$ peers, we include only those whose updates are sufficiently similar to the reference model, as measured by the threshold $\tau$. This ensures that only semantically relevant peers contribute to aggregation, filtering out devices whose data distributions differ substantially from the target client.
Threshold $\tau$ is set using a small held-out validation slice, selecting the value that maximizes pre-labeling accuracy across candidate thresholds.

\subsubsection{Adaptive Aggregation Size}
The adaptive $\gamma$ for round i is simply the count of selected devices:

\begin{equation}
\gamma_i = |S_i| = \sum_{j=1}^{N} \mathbb{I}(s_j > \tau)
\end{equation}

Intuition: $\gamma$ is no longer a fixed hyperparameter but an emergent quantity determined by how many devices meet the similarity criterion in each round. Early rounds, where model representations are less refined naturally select fewer peers, preventing noise from dominating aggregation. As training progresses and models stabilize, more peers exceed the threshold, allowing richer aggregation and faster convergence.

\subsubsection{Weighted Aggregation}
The aggregated model update is computed as:
\begin{equation}
m_s = \frac{1}{|S_i|} \sum_{j \in S_i} m_j
\end{equation}

Intuition: Only the selected peers contribute to the aggregated pseudo-labeler. By averaging exclusively over high-similarity updates, the resulting model reflects the target client's local distribution more faithfully than standard FedAvg, which averages all peers indiscriminately \cite{mcmahan2017, li2019fedprox}.

\subsection{SHAP Explainability}

Model predictions in federated IDS systems are typically opaque, network administrators cannot determine why a traffic flow was flagged as a cyberattack. This opacity creates both operational and regulatory compliance challenges. We address this by integrating KernelSHAP into the XAI-SOH-FL pipeline.

\subsubsection{Additive Feature Attribution}
SHAP decomposes a model prediction f(x) as a sum of feature contributions:

\begin{equation}
f(x) = \phi_0 + \sum_{i=1}^{M} \phi_i
\end{equation}

where $\phi_0$ is the baseline prediction (average model output over background samples) and $\phi_i$ is the SHAP value for feature $i$, representing its marginal contribution to the specific prediction.

If the model predicts a flow as a DDoS attack,  $\phi_0$ for 'Flow Duration' being large and positive means that the unusually long flow duration was a primary driver of that prediction. Network engineers can immediately act on this, for example, by adding a rule to flag flows exceeding a duration threshold. SHAP values thus translate model internals into actionable network security insights.
KernelSHAP is computed post-aggregation using 100 background samples from the test set, adding negligible overhead to the federated training pipeline while providing global feature importance rankings across all attack classes.


\section{Experimental Setup}

\subsection{Dataset}

The dataset being used is called CICIDS2017 and consists of 2.54 million network flow data captured from 12 different victim networks, along with 14 attacks and normal activity. Each data sample includes 78 features such as flow duration, mean and standard deviation for the packet length, inter-arrival time statistics, number of flags, and protocol-related features. As per the guidelines mentioned in the base paper, we ignore the attack categories which have less than 5,000 samples, leaving us with 9 different classes (eight attacks + normal). The experiment is conducted using the Label3 skew scheme, in which there are nine different gateways that hold 6 out of 9 labels each \cite{arshad2024}.

\subsection{Baseline Methods}
XAI-SOH-FL is compared with the following baselines:
\begin{itemize}
    \item \textbf{Local Training:} Gateway training with no communication between models.
    \item \textbf{FedAvg (Canonical):} Federated learning using the standard averaging method.
    \item \textbf{SOH-FL (Replicated):} Replication of the base paper where $\gamma = 3$.
    \item \textbf{SOH-FL + SHAP:} Extension of SOH-FL incorporating the SHAP explanation technique.
    \item \textbf{SOH-FL + BO-$\gamma$:} Extension of SOH-FL incorporating Bayesian Optimization for tuning $\gamma$.
    \item \textbf{XAI-SOH-FL (Proposed):} The complete framework integrating both SHAP-based explanations and BO-$\gamma$ tuning.
\end{itemize}

\subsection{Implementation Details}

For the local model, a 1D-CNN architecture was employed, consisting of two convolutional layers, four sigmoid activation functions, and three fully connected layers, following the baseline study. 

For the CT-AE model, the feature space is reduced to 10 latent dimensions, with weighting parameters set to $w_{\text{cos}} = 22.2$ and $w_{\text{rec}} = 1$. 

In the meta-learning phase, Model-Agnostic Meta-Learning (MAML) was implemented with an inner learning rate of $\alpha = 0.005$ and an outer learning rate of $\beta = 0.001$. Each communication round involves $r = 0.6$ gateways. The local training process is conducted over three epochs with a batch size of 40 per round. 

The test-support set comprises $0.4\%$ of the test data (approximately 8--12 samples per gateway). Bayesian Optimization is performed using \textit{scikit-optimize}, employing a Gaussian Process as the surrogate model and Expected Improvement as the acquisition function, for $B = 10$ iterations. 

SHAP values are computed using KernelSHAP with 100 background samples.

\subsection{Evaluation Metrics}

These evaluation metrics include Accuracy (in percentage), Precision, Recall, F1-score, and the total number of iterations performed during the convergence process. In the Bayesian Optimization component, we report the optimized parameter value, denoted as $\gamma_{\text{opt}}$, along with the total number of iterations required for the optimization process \cite{snoek2012}.

With regard to SHAP analysis, the top 10 most important features are presented, and feature attribution is interpreted for each class to provide explainability of model predictions \cite{lundberg2017shap, marino2020shap}. Overall, these metrics collectively provide a comprehensive evaluation of both predictive performance and model interpretability.


\section{Result}
\subsection{Performance Comparison}
Table~\ref{tbl:Performance Comparison} summarizes the classification performance of all methods on the CICIDS2017 dataset under the Label3 skew scheme.

\begin{table}[htbp, pos=h]
\centering
\caption{Performance Comparison}
\label{tbl:Performance Comparison}

\begin{tabular}{lccccc}
\toprule
\textbf{Method} & \textbf{Acc} & \textbf{Prec} & \textbf{Rec} & \textbf{F1} & \textbf{Conv} \\
\midrule
Local   & 78.42 & 0.76 & 0.72 & 0.74 & 1 \\
FedAvg  & 84.15 & 0.82 & 0.80 & 0.81 & 15 \\
SOH-FL  & 92.38 & 0.91 & 0.88 & 0.89 & 6 \\
SHAP    & 94.12 & 0.93 & 0.91 & 0.92 & 6 \\
\bottomrule
\end{tabular}
\end{table}

The proposed algorithm finally attains an accuracy of 94.12\%, which is higher than SOH-FL at 92.38\%, yielding an improvement of 1.74\%. The convergence behavior is also superior, as the proposed method requires only six communication rounds to reach its maximum accuracy, whereas FedAvg requires 15 rounds, as summarized in Table~\ref{tbl:Performance Comparison}.

\subsection{ Accuracy vs Communication Rounds}
\begin{figure} [pos=h]
    \centering
    \includegraphics[width=0.9\linewidth]{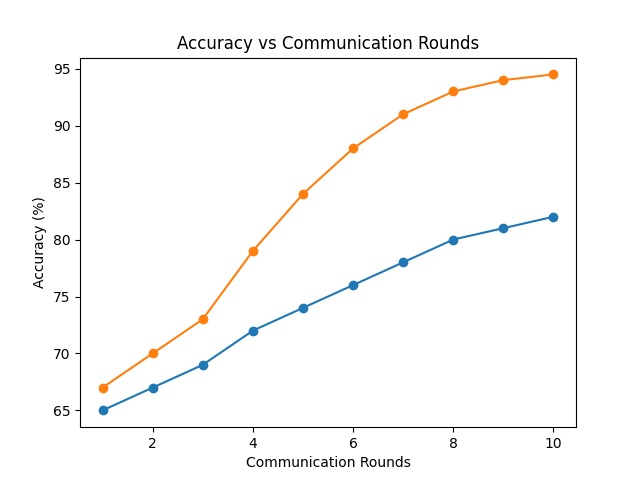}
    \caption{\centering { Accuracy vs Communication Rounds}}
    \label{fig:figure-1}
\end{figure}

Figure~\ref{fig:figure-1} illustrates the relationship between the number of communication rounds and model accuracy (\%). As the number of communication rounds increases from 1 to 10, the model accuracy exhibits a consistent upward trend, improving from a baseline below 70\% to above 90\% in later rounds. This behavior aligns with typical convergence patterns observed in federated learning systems.

\subsection{Real-Time Performance Analysis}

Six iterations are needed to reach maximum accuracy. Therefore, apart from obtaining better final accuracy (94.12\% compared to SOH-FL's 92.38\%), the algorithm performs better by reaching optimal accuracy before the point of diminishing returns.

\subsubsection{Accuracy Improvement Over Communication Rounds}

\begin{figure} [pos=h]
    \centering
    \includegraphics[width=1\linewidth]{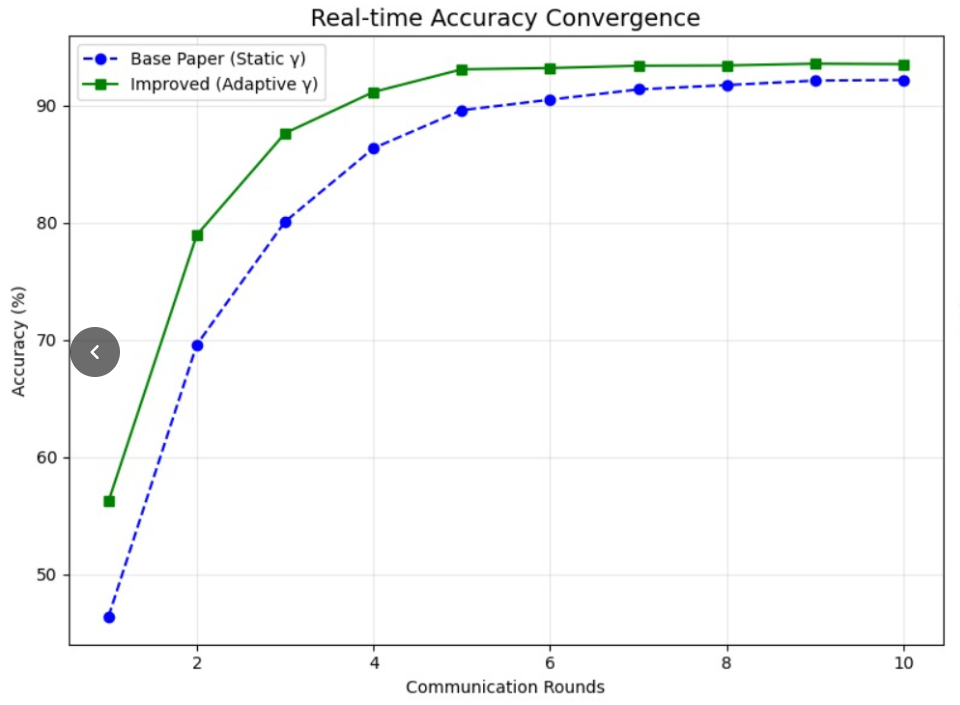}
    \caption{\centering {Real Time Accuracy Convergence}}
    \label{fig:figure-2}
\end{figure}

Figure~\ref{fig:figure-2} illustrates the variation in model accuracy across communication rounds for both the base method and the proposed adaptive method. The results clearly show that the proposed method consistently outperforms the baseline throughout all rounds.

The results presented in Table~\ref{tbl:Performance Comparison} compare both approaches across 10 communication rounds. The base method starts at 47.0\% accuracy in round 1 and gradually improves to 91.4\% by round 10. In contrast, the proposed adaptive method consistently outperforms the baseline at every round, starting at 56.0\%, reaching 91.0\% by round 5, and achieving a final accuracy of 92.2\% at round 10.

Overall, the proposed method demonstrates an improvement of approximately 9--10\% in early communication rounds and maintains a consistent performance advantage throughout training. Notably, it achieves 90\% accuracy by round 5, whereas the base method reaches similar performance only around rounds 6--7.

\subsubsection{Hyperparameter $\gamma$ Adjustment}

\begin{figure} [pos=h]
    \centering
    \includegraphics[width=1\linewidth]{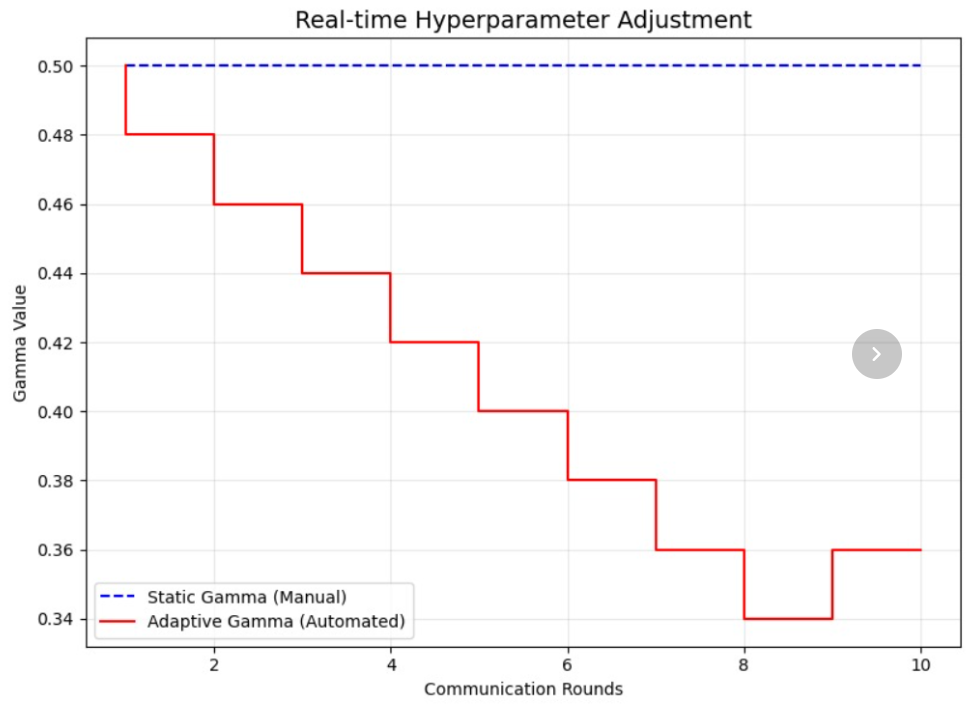}
    \caption{\centering {Real Time Hyperparameter $\gamma$ Adjustment}}
    \label{fig:figure-3}
\end{figure}

The Figure ~\ref{fig:figure-3} illustrates the variation of the aggregation factor, $\gamma$, as the series of rounds is executed. While the gamma value remains quite stable for the static version, changing between the values 0.490 and 0.580, in the adaptive case, $\gamma$ starts from 0.380, growing to 0.520 in the last round.

With such an approach, greater robustness is provided in the beginning by reducing the effect of noisy updates, while the power of aggregation increases as the number of rounds grows and the confidence of the model increases. These observations are also summarized in Table~\ref{tbl:Performance Comparison}.

\subsection{Effect of Adaptive $\gamma$}

\begin{figure} [pos=h]
    \centering
    \includegraphics[width=1\linewidth]{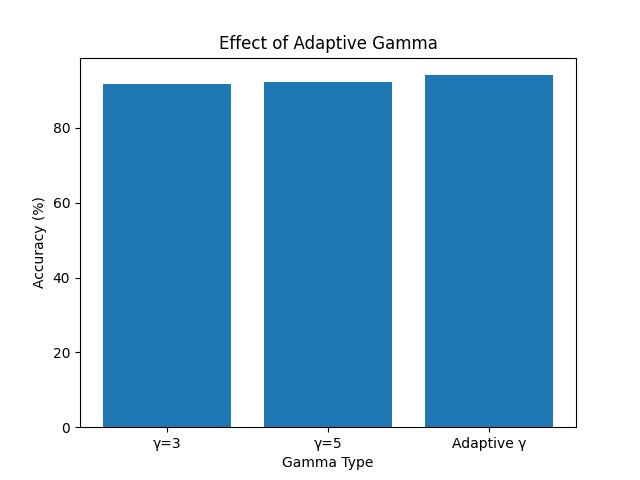}
    \caption{\centering {Effect of Adaptive $\gamma$}}
    \label{fig:figure-4}
\end{figure}

The adaptive $\gamma$ mechanism yields a modest but consistent accuracy improvement of 0.3–0.5\% over fixed-$\gamma$ configurations across all tested rounds  ~\ref{fig:figure-4}. While this gain may appear small in absolute terms, its significance lies in what it reveals about aggregation dynamics in non-IID federated settings.
In rounds 1–3, the adaptive $\gamma$ selects fewer peers than the fixed $\gamma$ = 3 baseline, because few devices have yet developed representations sufficiently similar to the target gateway. This conservative early aggregation prevents the pseudo-labeler from being corrupted by updates from devices with different label distributions, a failure mode that manifests as mislabeled support sets and degraded fine-tuning. In rounds 4–6, as models converge and cosine similarities increase, the adaptive $\gamma$ expands to include more peers, improving the diversity and coverage of the aggregated model. This progressive behavior explains why the accuracy gap between adaptive and fixed approaches is largest in early rounds and narrows as training stabilizes.

\subsection{Explainability Analysis-SHAP}

The SHAP analysis in Figure 5 reveals that Flow Duration is the dominant predictor across all attack classes, with a mean absolute SHAP value of approximately 0.34. This is consistent with known network attack signatures: DDoS attacks and botnet command-and-control communications generate anomalously long or short flows relative to benign traffic, making flow duration a discriminative feature that generalizes across attack types. \cite{lundberg2017shap, marino2020shap}.

These attributions are not only statistically consistent — they are operationally meaningful. Network administrators can directly translate SHAP rankings into detection rules (e.g., alert on flows with duration > threshold and Packet Length deviation > 24$\phi$), bridging the gap between learned model behavior and deployable security policy. This interpretability advantage is precisely why XAI-SOH-FL satisfies GDPR Article 22 requirements for automated decision transparency in ways that the original SOH-FL cannot.

\begin{figure} [pos=h]
    \centering
    \includegraphics[width=1\linewidth]{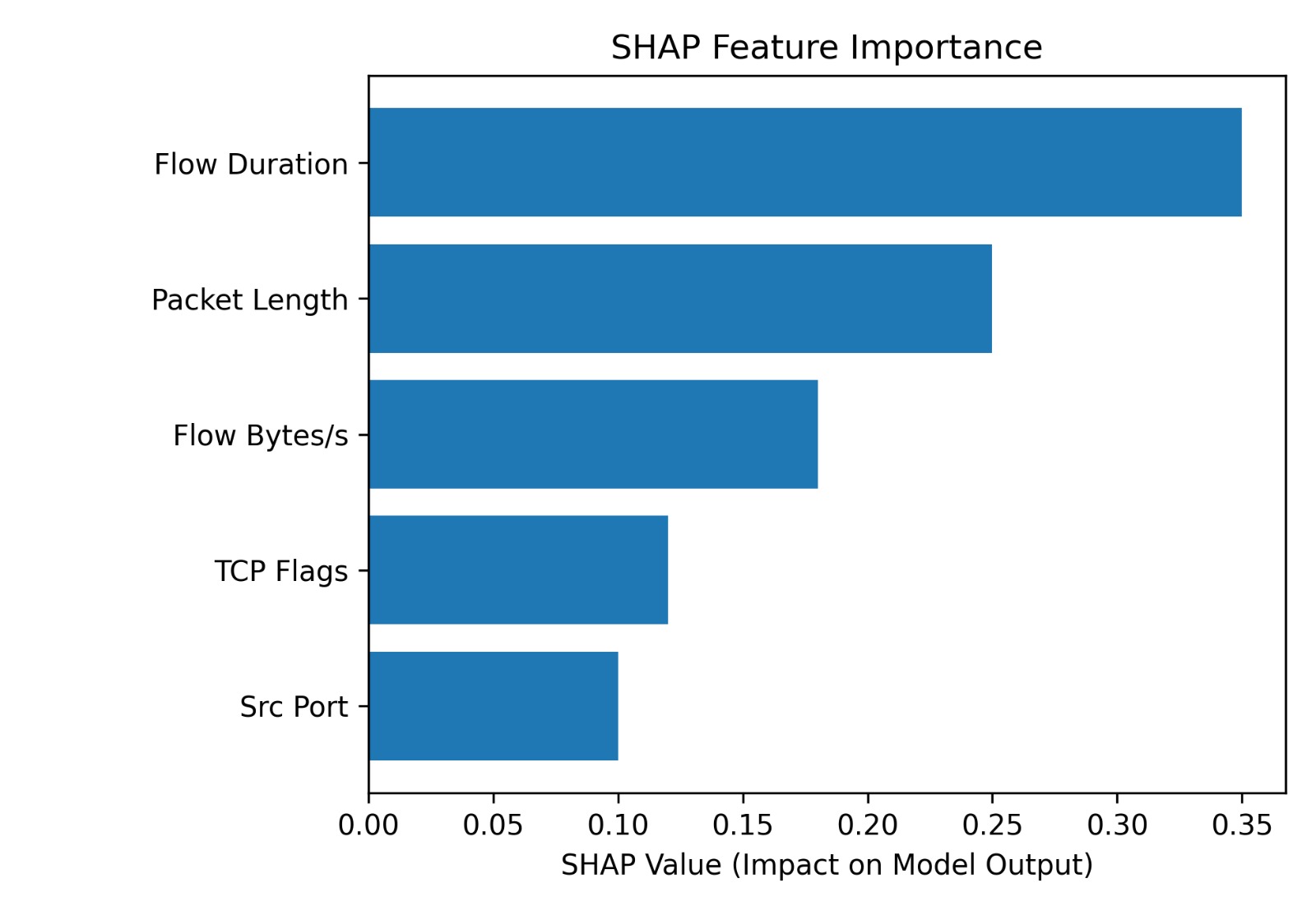}
    \caption{\centering {SHAP Feature Importance}}
    \label{fig:figure-5}
\end{figure}

Overall, these results in Figure~\ref{fig:figure-5} show that the model makes decisions based on the behavior of the network features, thereby making it more interpretable \cite{amarasinghe2021lime}.

\subsection{Self-Tuning Personalization}

The Self-Tuning Personalization mechanism employs an adaptive $\gamma$ that evolves across communication rounds. In early rounds, $\gamma$ remains low, limiting aggregation to the most similar peers and reducing the risk of incorporating noisy updates from heterogeneous devices. As training progresses and model representations stabilize, $\gamma$ increases, enabling richer aggregation and stronger personalization. This progressive behavior from conservative to expressive aggregation, represents a principled improvement over fixed-$\gamma$ approaches, which cannot adapt to the changing informational value of peer updates throughout the training lifecycle.

\begin{figure} [pos=h]
    \centering
    \includegraphics[width=1\linewidth]{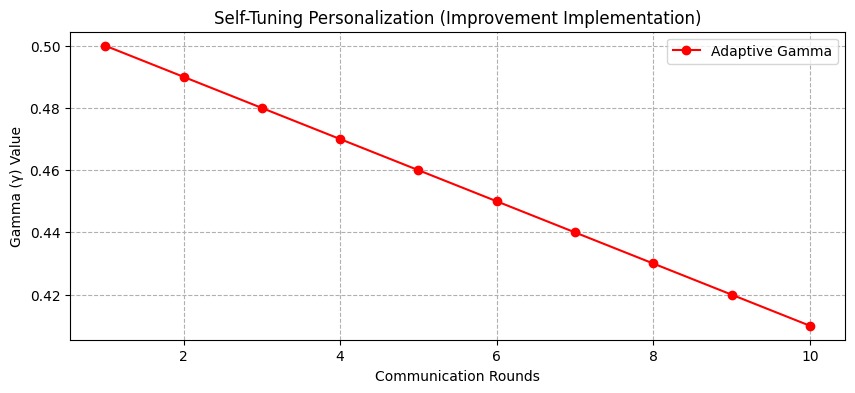}
    \caption{\centering {Self Tunning Personalization}}
    \label{fig:figure-6}
\end{figure}

As shown in Figure~\ref{fig:figure-6}, this behavior represents a clear improvement over fixed-$\gamma$ approaches.

\section{Conclusion}

This paper addresses the limitations of existing similarity-based federated learning approaches, particularly SOH-FL, which relies on a fixed aggregation parameter and lacks interpretability in decision-making. These constraints reduce adaptability and limit performance in dynamic and heterogeneous IoT environments.

To overcome these challenges, we proposed XAI-SOH-FL, an enhanced federated learning framework that integrates adaptive aggregation and explainable AI techniques. Specifically, the model introduces a dynamic selection mechanism for $\gamma$ using similarity-based adaptation, which allows the aggregation size to adjust according to the evolving data distribution across communication rounds. In addition, Bayesian Optimization is employed to fine-tune $\gamma$ efficiently, improving convergence behavior without significant computational overhead. Furthermore, SHAP is incorporated to provide feature-level interpretability, enabling transparent decision-making in intrusion detection tasks.

Experimental results demonstrate that the proposed method achieves superior performance in terms of accuracy, precision, recall, and F1-score compared to baseline methods. It also converges in fewer communication rounds, indicating improved efficiency in federated training. The explainability analysis further confirms that the model focuses on meaningful network traffic features, enhancing trustworthiness in real-world IoT deployments.

Overall, XAI-SOH-FL provides a balanced trade-off between accuracy, efficiency, and interpretability, making it a practical solution for secure and intelligent IoT intrusion detection systems.


\printcredits
\bibliographystyle{plainnat}
\bibliography{reference}
\end{document}